\documentclass[prd,twocolumn,superscriptaddress,floatfix,amsmath,amssymb,amsfonts,longbibliography]{revtex4-2}

\usepackage{float} 
\usepackage{scalerel}
\usepackage[normalem]{ulem}
\usepackage[english]{babel}
\usepackage{graphicx}
\usepackage{dcolumn}
\usepackage{bm}
\usepackage{blindtext}
\usepackage{verbatim}
\usepackage{relsize}
\usepackage{mathrsfs}
\usepackage{musicography}
\usepackage{amsmath}
\usepackage{blindtext}
\usepackage{cancel}
\usepackage{physics}
\usepackage{epstopdf}
\usepackage{mathtools}
\usepackage{blindtext}
\usepackage{tensor}
\usepackage{xcolor}
\usepackage[usenames,dvipsnames]{pstricks}
\usepackage{epsfig}
\usepackage{pst-grad}
\usepackage{pst-plot} 
\usepackage{hyperref}
\usepackage{verbatim}
\usepackage{slashed}
\usepackage{dsfont}
\usepackage[english]{babel}
\usepackage{amsmath,amssymb,amsthm} 
\usepackage{lipsum}
\usepackage{stmaryrd}
\usepackage{trimclip}

\makeatletter
\DeclareRobustCommand{\shortto}{%
  \mathpalette\short@to\relax%
}

\newcommand{\short@to}[2]{%
  \mkern2mu
  \clipbox{{0.3\width} 0 0 0}{$\m@th#1\vphantom{+}{\shortrightarrow}$}%
  }

\newcommand{\ii}{\mathrm{i}}

\renewcommand{\H}{\mathscr{H}}

\newcommand{\s}{\hat{\sigma}}
\newcommand{\tc}[1]{\textsc{#1}}

\newcommand{\faketwo}[1]{
}

\newcommand{\secl}[1]{%
\lettersec{#1}}

\newcommand{\lettersec}[1]{\noindent\textbf{\textit{#1---}}}

\theoremstyle{plain}

\theoremstyle{definition}

\allowdisplaybreaks[1] 

\begin{document}

\title{Subsystem entanglement and separability in quantum reference frames}


\author{T. Rick Perche}
\email{rick.perche@su.se}

\affiliation{Nordita,
KTH Royal Institute of Technology and Stockholm University,
Hannes Alfvéns väg 12, 23, SE-106 91 Stockholm, Sweden}

\author{Natália Salomé Móller}
\email{natalia.moller@savba.sk}

\affiliation{Research Center for Quantum Information, Institute of Physics, Slovak Academy of Sciences, Dúbravská Cesta 9,
84511 Bratislava, Slovakia}

\author{Guilherme Franzmann}
\email{guilherme.franzmann@su.se}

\affiliation{Nordita,
KTH Royal Institute of Technology and Stockholm University,
Hannes Alfvéns väg 12, 23, SE-106 91 Stockholm, Sweden}

\affiliation{Department of Philosophy, Stockholm University, Stockholm, Sweden}

\affiliation{Basic Research Community for Physics e.V., Mariannenstraße 89, Leipzig, Germany}

\begin{abstract}
    We find a necessary condition for subsystems to become entangled after a quantum reference frame transformation. We then distinguish between subsystems that admit separable descriptions relative to a quantum reference frame, and those that do not. On the one hand, we show that separable descriptions maximize the entanglement internal to the subsystem, and relate our results with the conservation of entanglement and coherence under quantum reference frame transformations. On the other hand, systems that do not admit a separable description relative to any subsystem display a form of entanglement that is genuine to the quantum reference frame formalism.
\end{abstract}

\maketitle


{\secl{Introduction}\label{sec:intro}The notion of a quantum reference frame (QRF) extends the relational perspective of classical physics into quantum mechanics, enabling different quantum systems to provide distinct descriptions of the same physical situation~\cite{Aharonov:1984zz,Bartlett:2006tzx,GiacominiQRF2019,ac,Vanrietvelde_2020,H_hn_2021,H_hn_2022,acNeutral2021,castroruiz2023relativesubsystemsquantumreference,Miyadera_2016,Loveridge_2017,Loveridge_2018,głowacki2024quantumreferenceframeshomogeneous,JanQRFHomog2024,castroruiz2025interpretingquantumreferenceframe}. QRF transformations often lead to striking effects such as changes in subsystem structure \cite{AliAhmad:2021adn}, in the operational meaning of observables~\cite{JanQRFHomog2024,JanOperational2025}, frame-dependent superposition, and the appearance or disappearance of entanglement~\cite{GiacominiQRF2019,Cepollaro:2024rss}. As this perspective also provides a path towards quantum gravity by extending the equivalence principle to the quantum domain~\cite{GiacominiQRF2019,Giacomini_2022,giacomini2023einsteinsequivalenceprinciplesuperpositions}, it is essential to classify under which conditions these effects appear, and which of these are due to intrinsic features of quantum reference frames.

Here, we focus on the frame dependence of entanglement, which immediately raises a basic question: when can a change of QRF generate entanglement within a collection of systems that is not itself used as either the initial or final frame? As we will see, the form of a QRF transformation alone answers neither of these questions by itself. Rather, we show that entanglement can appear within the collection only if, in the initial description, there already exists entanglement across the collection and the other subsystems—even when the reduced state of the collection is separable. This result leads to a second question: whether correlations between the collection and the remaining frames can be removed by an appropriate choice of reference frame. 

Separability plays a central role in addressing this question. If there exists a QRF description in which the collection factorizes from the remaining subsystems, then it can be assigned an independent state in that description. This by itself doesn't require the systems within the collection to be mutually separable: entanglement internal to the collection is distinct from entanglement between the collection and the other reference frames. In fact, we show that a separable description maximizes the entanglement internal to the collection, since changing away from it acts as a random local-unitary channel and cannot increase entanglement. 

Moreover, the absence of a separable description identifies physical situations that go beyond the typical use of QRFs: if no available QRF renders the collection separable from the remaining subsystems, we show that its nonseparability cannot be removed by a change of perspective. This marks a genuinely quantum phenomenon specific to the QRF framework. As we discuss, the existence of a separable description thus provides a sharp criterion for distinguishing correlations that can be transformed away from those that are irreducible under changes of quantum reference frames.

In this Letter, we derive the necessary condition underlying this distinction and use it to characterize collections of subsystems that admit separable descriptions relative to a QRF. We also relate our results to the conservation of coherence and entanglement under QRF transformations proved in \cite{Cepollaro:2024rss}, and introduce a frame-independent quantifier that detects when no separable description exists.}

\smallskip

\secl{Quantum reference frames}\label{sec:QRFs}Quantum reference frames are often introduced with a relational motivation, where one wishes to describe a set of $N$ physical systems relative to one of its constituents. Following the description of~\cite{ac}, consider a group $G$ and a set of $N$ systems that can each be described in a Hilbert space $\H_i \cong L^2(G)$. Within each $L^2(G)$, it is possible to write an orthonormal basis~\cite{f1} labeled by group elements, $\{\ket{g}\}_{g\in G}$, where we define the action of left multiplication in the group as a unitary operation: $\ket{g'g} = \hat{U}(g')\ket{g}$. The identity of the group can then be used to label a ``classical reference frame'' (such as the origin in $\mathbb{R}^3$~\cite{GiacominiQRF2019}), and the remaining frame states are obtained by group multiplication. We denote by $\dd g$ the left-invariant measure of the group satisfying
\begin{equation}
    \openone = \int \dd g \ket{g}\!\!\bra{g},
\end{equation}
so that $\langle g|g'\rangle = \delta(g^{-1}g')$, where $\delta(g)$ denotes the Dirac delta with respect to the measure $\dd g$ centered at the group origin~\cite{f2}. Within these assumptions, each system described by $L^2(G)$ can be thought of as defining a reference frame. 

In this context, if $e$ is the identity of the group, we say that a state of the form $\ket{\Psi}^{(1)} = \ket{e}_{{1}}\otimes \ket{\psi_\text{rel}}^{(1)}$ describes the systems in $\mathscr{H} = \mathscr{H}_1\otimes...\otimes \mathscr{H}_N \cong L^2(G)^{\otimes N}$ with respect to system $1$. If one wishes to perform a change of reference frame such that the system is instead described with respect to the second factor of $\H$, we can perform a change of QRF by applying the unitary transformation
\begin{equation}\label{eq:S12}
    \hat{S}^{1\shortto 2} = \hat{\Pi}_{12} \int \dd g \,\,\openone\otimes \ket{g^{-1}}\!\bra{g}\otimes \hat{U}^\dagger(g)^{\otimes N},
\end{equation}
where $\hat{\Pi}_{12}$ is the swap operation acting on systems 1 and 2. Notice that applying $\hat{S}^{1\shortto 2}$ to a state $\ket{\Psi}^{(1)} = \ket{e}_1\otimes \ket{\psi_\text{rel}}^{(1)}$ results in a state $\ket{\Psi}^{(2)} = \hat{S}^{1\shortto 2}\ket{\Psi}^{(1)} = \ket{e}_2\otimes |{\psi}_\text{rel}\rangle^{(2)}$, changing the relative state $\ket{\psi_\text{rel}}^{(1)}$ to $|{\psi}_\text{rel}\rangle^{(2)}$.

\begin{figure}[h]
    \centering
    \includegraphics[width=7.3cm]{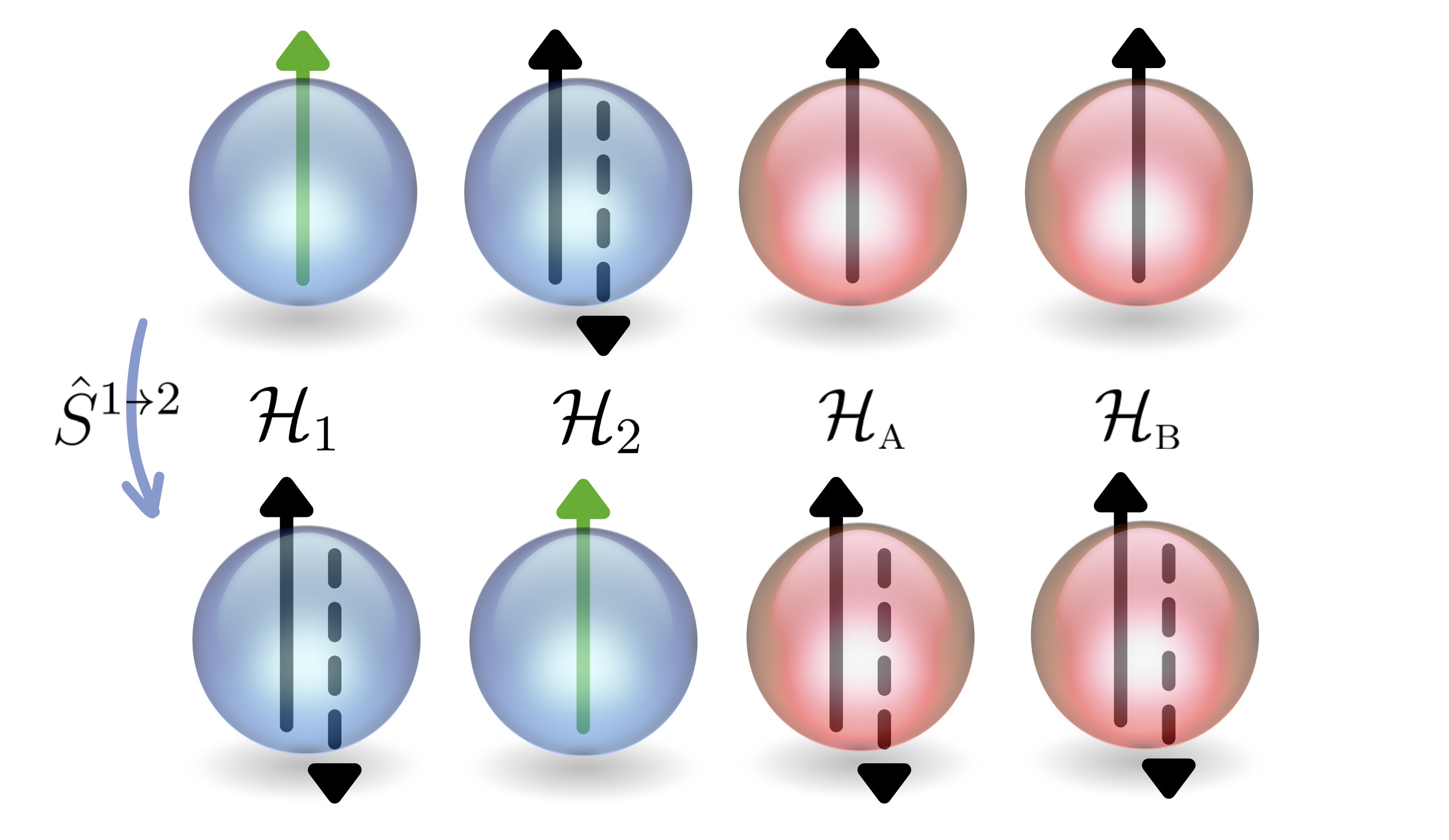}
    \caption{The figure depicts a change of reference frames from system 1 to system 2 (blue). The state of the current reference frame is denoted as a green arrow, and we explicitly identify the remaining subsystem AB (red) not used as references. 
    }
    \label{fig:L}
\end{figure}

A simple example for QRF transformations is when we consider qubits, each described in $\mathbb{C}^2 \cong L^2(\mathbb{Z}_2)$, with the symmetry group $\mathbb{Z}_2 = \{0,1\}$, represented by $0\mapsto\openone$ and $1\mapsto \s_x$. To illustrate the changes of entanglement structure under QRF transformations, we consider two examples in a system of four qubits, two that we will use as reference frames (1 and 2) and two systems (A and B) that we choose not to explicitly use as references. In this case, the change of reference frame unitary can be written as a CNOT gate of the form
\begin{equation}
    \hat{S}^{1\shortto2} = \hat{\Pi}_{12}\,\openone_1\otimes(\ket{0}\!\!\bra{0}_2\otimes \openone_{\tc{ab}} + \ket{1}\!\!\bra{1}_2\otimes \hat{\sigma}^x_\tc{a}\otimes \hat{\sigma}^x_\tc{b}).
\end{equation}

We motivate our initial question, of when entanglement between subsystems can be created through a QRF transformation by looking at two examples, inspired by~\cite{hohnQRelSub2023}.

{\noindent\textbf{Example 1.} \itshape 
Consider the case where the system of four qubits is described relative to the first qubit by the state
\begin{equation}\label{eq:psiEx1}
    \ket{\Psi}^{(1)} = \ket{0}_1\otimes \ket{+}_2\otimes\ket{00}_\tc{ab}.
\end{equation} 
We then perform the change of quantum reference frame to the second qubit, obtaining the state (described relative to qubit 2): 
\begin{align}
    \ket{\Psi}^{(2)} &= \hat{S}^{1\shortto2}\ket{\Psi}^{(1)} =  \ket{0}_2\otimes\tfrac{1}{\sqrt{2}}(\ket{000}_{1\tc{ab}}+\ket{111}_{1\tc{ab}})\nonumber\\
    &\equiv\ket{0}_2\otimes \ket{GHZ}_{1\tc{ab}}.
\end{align} 
Notice that the second component of the state above describes the first qubit. 

As discussed in~\cite{Cepollaro:2024rss}, under the change of reference frames, the coherences of the qubit $\ket{+}$ are converted into entanglement between the previous reference frame and the remaining qubits, mapping a superposition within one Hilbert space to a superposition in multiple Hilbert spaces, corresponding to an entangled state. In this example, the previous reference frame is necessary to witness the entanglement: the reduced state of the systems A and B (not involved in the reference frame transformation) remains separable: $\hat{\rho}_{\tc{ab}}^{(2)} = \tfrac{1}{2}(\ket{00}\!\!\bra{00} + \ket{11} \!\!\bra{11})$. Thus, no entanglement is created between the systems that are not involved in the QRF transformation.}

Notice that a change of reference frames can also produce an entangled state in AB. We illustrate this in the following example.

\smallskip
{\noindent\textbf{Example 2.} \itshape  Consider the case in which the initial state from the perspective of the reference system 1 is
\begin{equation}\label{eq:psiEx2}
    \ket{\Psi}^{(1)} = \frac{1}{\sqrt{2}}\ket{0}_1\otimes (\ket{0}_2\otimes\ket{\Phi_{+\ii}}_\tc{ab}+\ii \ket{1}_2\otimes\ket{\Phi_{-\ii}}_\tc{ab}),
\end{equation}
where $\ket{\Phi_\alpha} = \tfrac{1}{\sqrt{2}}(\ket{00} + \alpha \ket{11})$, with $|\alpha| = 1$. In this case, the reduced state of systems A and B is the separable state $\hat{\rho}_{\tc{ab}}^{(1)} = \tfrac{1}{2}(\ket{00}\!\!\bra{00} + \ket{11}\!\!\bra{11})$. When one transforms into the reference frame of system 2, the state then becomes
\begin{equation}\label{eq:psiEx22}
    \ket{\Psi}^{(2)} = \ket{0}_2\otimes \ket{+}_1\otimes \ket{\Phi_{+\ii}}_{\tc{ab}},
\end{equation}
which has an entangled partial state for systems A and B. Thus, a QRF transformation can create entanglement between the systems that are not the new and previous reference frames.} 

One could attempt to draw conclusions about entanglement generation from Examples 1-2 by noticing that the QRF transformation~\eqref{eq:S12} is a controlled unitary, thus generating entanglement only when the control is in a superposition. If the control system is in a coherent superposition of states corresponding to group elements, then the QRF transformation implements a conditional unitary that can entangle the control with whatever system it acts on. However, the crucial point for us is that this entanglement need not appear \textit{between the subsystems being described with respect to the previous and new QRFs}. As Example 1 illustrates, the controlled action of the QRF map can generate entanglement only between the old reference frame and the rest, while leaving the subsystem AB separable; only if the relative state of the reference frames is already correlated with AB (as in Example 2) can the same mechanism generate entanglement \emph{within} the AB subsystem.

\secl{Subsystem entanglement in QRFs}\label{sec:no-entanglement}We will now show a necessary condition for a given set of subsystems not involved in a quantum reference frame transformation (such as subsystems A and B in the examples above) to be entangled after a QRF transformation. Consider a system of $N$ quantum systems in $L^2(G)^{\otimes N}$ that is at first described with respect to the frame of system $k\in\{1,...,N\}$. We will consider a change of reference frame from system $k$ to system $l$ and study whether the entanglement within a subset $I \subset \{1,...,N\}$ changes under this transformation, with $l,k\in I^c$. 
Without loss of generality, we will assume that the ordering of Hilbert spaces is such that $I^c = \{1,...,m\}$ and $I = \{m+1,...,N\}$.

A general $N$-partite pure state $\ket{\Psi}$ described with respect to a given reference frame $k\in I^c$ can be written as
\begin{equation}\label{eq:geneerapsik}
    \ket{\Psi}^{(k)} = \int \dd \bm g \, \delta(g_k) f(\bm g) \ket{\bm g}_{I^c} \otimes \ket{\psi(\bm g)}_I,
\end{equation}
where $\bm g = (g_1,...,g_m)\in G^m$ and $\ket{\bm g}_{I^c} = \ket{g_1...g_m}$ describes a basis for the systems $I^c$ and the Dirac delta ensures that the reduced state of the frame $k$ is $\ket{e}_k$. When applied to a system described with respect to system $k$, the change of reference frame transformation to a system $l\in I^c$ takes the relatively simple form (see Appendix~\ref{app:generalPQRF})
\begin{equation}
    \hat{T}^{k\shortto l}  = \int \dd \bm g \,\delta(g_k)\,g_l^{-1}\!\cdot \ket{\bm g}\!\!\bra{\bm g}_{I^c} \otimes \hat{\mathcal{U}}^\dagger(g_l),
\end{equation}
where $\hat{\mathcal{U}}^\dagger(g_l) = \hat{U}_{m+1}^\dagger(g_l)...\hat{U}_N^\dagger(g_l)$ is the total separable unitary action in the subsystems in $I$ and $g_l^{-1}\cdot \ket{\bm g} = \ket{g_l^{-1}\bm g}$ denotes the left unitary action in the systems $1,...,m$. In Appendix~\ref{app:generalPQRF}, we show that a change of reference frames from $k$ to $l$ results in the state
\begin{equation}\label{eq:generalpsil}
    \ket{\Psi}^{(l)} = \int \dd \bm g \,\delta(g_l) \tilde{f}(\bm g) \ket{\bm g}_{I^c} \otimes |{\tilde{\psi}(\bm g)}\rangle_I,
\end{equation}
where $|{\tilde{\psi}(\bm g)}\rangle = \hat{\mathcal{U}}^{\dagger}(g_k^{-1})|\psi(g_k^{-1}\bm g)\rangle$ and $\tilde{f}(\bm g) = f(g_k^{-1}\bm g)$. This allows us to formulate the following theorem: 

\smallskip
{\noindent\textbf{Theorem.} \itshape Given a collection of subsystems $I$ and $k,l\notin I$,  if $\ket{\bm g}\!\!\bra{\bm g}_{I^c}\!\ket{\Psi}^{(k)}$ is separable for every $\bm g\in G^m$ from the perspective of a given reference frame $k$, then $\ket{\bm g}\!\!\bra{\bm g}_{I^c}\!\ket{\Psi}^{(l)}$ is separable for every $\bm g\in G^m$ in any other reference frame $l$.} 

\smallskip
The theorem follows from the fact that the states $|{\psi}(\bm g)\rangle_I$ in Eq.~\eqref{eq:geneerapsik} are related to $|\tilde{\psi}(\bm g)\rangle_I$ in Eq.~\eqref{eq:generalpsil} by local unitaries. We can use this result to specify under which conditions subsystems can be entangled after a QRF transformation between other subsystems. Notice that for the state $\ket{\Psi}^{(k)}$ in Eq.~\eqref{eq:geneerapsik}, the partial state of the subsystem in $I$ is of the form
\begin{equation}
    \hat{\rho}_I^{(k)} = \tr_{{I^c}}(\ket{\Psi}\!\!\bra{\Psi}^{(k)}) = \int \dd \bm g \delta(g_k)|f(\bm g)|^2 \ket{\psi(\bm g)}\!\!\bra{\psi(\bm g)}_I.
\end{equation}
If $\ket{\psi(\bm g)}_I$ is separable for all $\bm g \in G^m$ in Eq.~\eqref{eq:geneerapsik}, we conclude that the state $\hat{\rho}^{(k)}_I$ is not entangled, and there is no entanglement between the subsystems in $I$. This defines a sufficient condition for the absence of entanglement between subsystems after a QRF transformation: if $\ket{\bm g}\!\!\bra{\bm g}_{I^c}\!\ket{\Psi}^{(k)}$ is separable for every $\bm g\in G^m$ for a given reference frame $k$, $\ket{\Psi}^{(l)}$ does not contain entanglement within $I$ for any other reference frame $\mathscr{H}_l$ with $k,l\in I^c$. The contrapositive of this statement can be phrased as the following corollary:

\smallskip
{\noindent\textbf{Corollary.} \itshape If $\ket{\Psi}^{(k)}$ has entanglement within the subsystems in $I$, then for every reference frame $l\notin I$ (including $l=k$), $\ket{\bm g}\!\!\bra{\bm g}_{I^c}\!\ket{\Psi}^{(l)}$ must be non-separable for at least one $\bm g\in G^m$. }

\smallskip

This result states that for there to be entanglement within $I$, at least for one combination of axes $\bm g\in G^m$ of the complementary systems in $I^c$, the systems in $I$ must be ``seen'' in an entangled state. Notice that the condition above is necessary but not sufficient. This can be seen in Example 2, where the state $\ket{\Psi}^{(1)}$ had no entanglement within $I$, but the relative states $\ket{00}\!\!\bra{00}_{12}\ket{\Psi}^{(1)}_{\tc{ab}}$, $\ket{01}\!\!\bra{01}_{12}\ket{\Psi}^{(1)}_{\tc{ab}}$, $\ket{00}\!\!\bra{00}_{21}\ket{\Psi}^{(2)}_{\tc{ab}}$, and $\ket{01}\!\!\bra{01}_{21}\ket{\Psi}^{(2)}_{\tc{ab}}$ were all entangled.

In Appendix~\ref{app:mixed}, we also derive a generalization of the entanglement condition for the case of mixed states. Specifically, we show that if $k,l\notin I$ and a mixed state $\hat{\rho}^{(k)}$ described with respect to reference frame $k$ has entanglement within $I$, then in any reference frame $l$, at least one of the operators $\ket{\bm g}\!\!\bra{\bm g}_{I^c} \hat{\rho}^{(l)} \ket{\bm g}\!\!\bra{\bm g}_{I^c}$ must be entangled for a given $\bm g \in G^m$.

\smallskip
\secl{QRF-separability}The characterization of changes in entanglement within subsystems under a QRF transformation presented in the previous section has an important application for the classification of systems that display behaviour intrinsic to the QRF formalism. In this section, we will define a reference frame independent notion of separability of subsystem states, from which a genuinely new QRF notion of entanglement arises.



If a collection of subsystems $I$ admits a separable description relative to a given system $k\in I^c$, that is, if
\begin{equation}\label{eq:CRFdesc}
    \ket{\Psi}^{(k)} = \ket{\phi}_{I^c}^{(k)}\otimes \ket{\psi}_I^{(k)},
\end{equation}
we will say that the state of $I$ is \textit{QRF-separable}. In this case, the reduced state of system $I$ relative to system $k$ can be described by the pure state $\ket{\psi}_I^{(k)}$ and is uncorrelated with the remaining subsystems in $I^c$. In this sense, a QRF-separable subsystem is separable up to a change of QRFs. 

The necessary conditions for entanglement proved in the last section allow us to conclude important facts about system $I$ when it admits a separable description. If $k$ is a separable description for system $I$ and $l\in I^c$, the fact that the reduced state of system $I$ after the QRF transformation to system $l$, $\hat{\rho}_I^{(l)}$ is obtained from $\ket{\psi}_I^{(k)}$ by a random local unitary channel (see Appendix~\ref{app:proof} for details) shows that QRF transformations cannot increase the entanglement within $I$ when compared to a separable description. That is, separable descriptions of system $I$ are also the descriptions that maximize the entanglement within $I$. In particular, the fact that $\ket{\bm g}\!\!\bra{\bm g}_{I^c}\ket{\Psi}^{(k)} = \braket{\bm g}{\phi}_{I^c}^{(k)} \ket{\bm g}_{I^c}\otimes \ket{\psi}_I^{(k)}$ (from Eq.~\eqref{eq:CRFdesc}), implies that if $\ket{\psi}_I^{(k)}$ is separable, QRF transformations cannot produce entanglement within $I$. 







Notice that a collection of subsystems $I$ described relative to frame $k$ does not admit a separable description with respect to any other subsystem if and only if the vectors $\hat{\mathcal{U}}^\dagger(g_l)\ket{\bm g}\!\!\bra{\bm g}_{I^c}\!\ket{\Psi}^{(k)}$ are not proportional for each $l\in I^c$. Conversely, if there exists a given system $l\in I^c$ where the vectors $\hat{\mathcal{U}}^\dagger(g_l)\ket{\bm g}\!\!\bra{\bm g}_{I^c}\!\ket{\Psi}^{(k)}$ are proportional for all $\bm g \in G^m$, then $I$ admits a separable description with respect to system $l$. 


The concept of QRF-separable subsystems naturally leads to the definition of QRF-entangled subsystems as those that do not admit a separable description with respect to any other subsystem. A QRF-entangled subsystem $I$ is then a subsystem that is entangled with the remaining systems in $I^c$ when described relative to any reference frame $k\in I^c$. In the following we display an example of a QRF-entangled subsystem:

\smallskip
{\noindent\textbf{Example 3.} \itshape 
Consider four qubits described relative to the first qubit by the state
\begin{equation}\label{eq:psiEx1}
    \ket{\Psi}^{(1)} = \ket{0}_1\otimes\frac{1}{\sqrt{2}}( \ket{0}_2\otimes\ket{0}_\tc{a}\otimes\ket{0}_\tc{b}+ \ket{1}_2\otimes\ket{1}_\tc{a}\otimes\ket{+}_\tc{b}).
\end{equation} 
Consider the subsystem $I$ composed of A and B, and notice that it does not admit a separable description with respect to system 1, or system 2. Indeed, after a change of QRF to system 2, we obtain
\begin{align}
    \ket{\Psi}^{(2)} &= \ket{0}_2\!\otimes\!\ket{0}_\tc{a}\!\!\otimes\!\frac{1}{\sqrt{2}}( \ket{0}_1\!\!\otimes\!\ket{0}_\tc{b}+ \ket{1}_1\!\!\otimes\!\ket{+}_\tc{b}),
\end{align} 
showing non-factorizability of system $I$ with respect to any available frame, thus showing this subsystem does not admit a separable description with respect to any of the available frames. We also note that subsystem A by itself admits a separable description with respect to system 2, and that if system A were not present, no subsystem would admit a separable description with respect to any other frame.}

Subsystems $I$ that do not admit a separable description with respect to any system in $I^c$ are precisely the subsystems that display bipartite entanglement with the remaining systems that is intrinsic to the QRF formalism. For pure states, this entanglement can be quantified through the minimal entanglement entropy (or, more generally, any faithful entanglement quantifier) of subsystem $I$ among all possible descriptions relative to systems $k\in I^c$:
\begin{equation}
    E_{\text{QRF}}(I) = \min_{k\in I^c}\left(S(\hat{\rho}_I^{(k)})\right),
\end{equation}
where $S(\hat{\rho}) = - \text{tr}(\hat{\rho} \log \hat{\rho})$ is the von Neumann entropy and $\hat{\rho}_I^{(k)} = \text{tr}_{I^c}(\ket{\Psi}\!\!\bra{\Psi}^{(k)})$ is the reduced state of the subsystems in $I$ described relative to system $k$. One then has $E_\text{QRF}(I) = 0$ if and only if the collection of subsystems in $I$ admits a separable description relative to a given subsystem $k\in I^c$.



\secl{Subsystem entanglement and coherence}In~\cite{Cepollaro:2024rss}, it was proven that, for pure states, the sum of entanglement and coherence is conserved under quantum reference frame transformations. This result follows from the fact that the set of diagonal elements in the group basis is preserved under QRF transformations. Indeed, the diagonal elements associated with the density operator $\hat{\rho}^{(k)} = \ket{\Psi}\!\!\bra{\Psi}^{(k)}$ from Eq.~\eqref{eq:geneerapsik} are
\begin{equation}\label{eq:diagk}
    D_{(k)} = \left\{|f(\bm g)_{g_k = e}|^2 |\braket{\psi(\bm g)}{\bm g_I}|^2:(\bm g,\bm g_I)\in G^N\right\},
\end{equation}
while, in the reference frame $l$, we have (from Eq.~\eqref{eq:generalpsil})
\begin{equation}\label{eq:diagl}
    D_{(l)} = \left\{|\tilde{f}(\bm g)_{g_l = e}|^2 |\langle{\tilde{\psi}(\bm g)}|{\bm g_I}\rangle|^2:(\bm g,\bm g_I)\in G^N\right\}.
\end{equation}
The relationship between $|\tilde{\psi}(\bm g)\rangle$ and $\ket{\psi(\bm g)}$ then implies $\langle{\tilde{\psi}(\bm g)}|{\bm g_I}\rangle = \langle\psi(g_k^{-1}\bm g)|g_k^{-1}\bm g_I\rangle$. Together with the fact that $\tilde{f}(\bm g) = f(g_k^{-1} \bm g)$ these imply that Eqs.~\eqref{eq:diagk} and~\eqref{eq:diagl} are identical with the reparametrization $\bm g \mapsto g_k^{-1}\bm g$, $\bm g_I \mapsto g_k^{-1}\bm g_I$. 

The fact that $D_{(k)} = D_{(l)}$ implies that the sum of entanglement and coherence is conserved, as this only depends on properties of the diagonal density matrix. More specifically, the results of~\cite{Cepollaro:2024rss} state that when changing reference frame from a system $k$ to a system $l$, we have
\begin{equation}\label{eq:sumEntCoh}
    \mathcal{E}_{l}\left(\ket{\Psi}^{(k)}\right) + \mathcal{C}_l\left(\ket{\Psi}^{(k)}\right) = \mathcal{E}_{k}\left(\ket{\Psi}^{(l)}\right) + \mathcal{C}_k\left(\ket{\Psi}^{(l)}\right),
\end{equation}
where $\mathcal{E}_l(\ket{\Psi}) = S(\hat{\rho}_l)$, with $S$ the von Neumann entropy and $\hat{\rho}_l = \Tr_{l^c}(\ket{\Psi}\!\!\bra{\Psi})$, is the entanglement entropy between system $l$ and the remaining systems, and $\mathcal{C}_l(\ket{\Psi}) = S(\Delta(\hat{\rho}_l))-S(\hat{\rho}_l)$, with $\Delta(\hat{\rho}_l) = \text{diag}(\hat{\rho}_l)$, is the relative entropy of coherence. The results of~\cite{Cepollaro:2024rss} then show that the sum of entanglement and coherence of the previous and new reference frames after a QRF transformation is conserved. 

Notice that the result of Eq.~\eqref{eq:sumEntCoh} also generalizes to the case where the states are mixed, however, in this case, the entanglement entropy is not an entanglement quantifier: in general, rather than entanglement plus coherence being conserved, what is conserved is the amount of information accessible by $l$ through group basis measurements (encoded in $\mathcal{C}_l(\ket{\Psi}^{(k)})$) added to the information inaccessible to it (encoded in $\mathcal{E}_{l}(\ket{\Psi}^{(k)})$).



{
We can now relate our results on separable descriptions to the conservation law of~\cite{Cepollaro:2024rss}. Consider a system described relative to subsystem $1$, with $I=\{3,\ldots,N\}$, and a QRF transformation to subsystem $2$. Assume that $I$ admits a separable description relative to system $1$, so that
\begin{equation}
    \ket{\Psi}^{(1)} = \ket{e}_1\otimes\ket{\varphi}_2\otimes\ket{\psi}_I^{(1)}.
\end{equation}
In this description, system $2$ is unentangled with $I$, and the entanglement within $I$ is maximal among all QRF descriptions obtained by changing reference frame within $I^c$. If the transformation from system $1$ to system $2$ is nontrivial (beyond local unitaries), then $\ket{\varphi}_2$ must contain coherence in the group basis. By the conservation law of~\cite{Cepollaro:2024rss}, this coherence is converted into entanglement involving the previous and new reference systems. Since $I$ started in a separable description, entanglement internal to $I$ cannot increase: the appearance of correlations with the previous frame implies that entanglement initially contained within $I$ is redistributed to the larger system $\{1\}\cup I$. This provides an operational interpretation of why separable descriptions maximize entanglement within subsystems: they minimize the entanglement with frames and remaining systems.
}

\smallskip



\secl{Conclusions}We established a necessary condition for entanglement to be present within a collection of subsystems after a QRF transformation, and a sufficient condition for no entanglement to be produced. In summary, a subsystem $I$ can only be entangled in some reference frame if the projected state $\ket{\bm g}\!\!\bra{\bm g}_{I^c}\ket{\Psi}$ is entangled for at least one $\bm g \in G^m$. This identifies the projected relative states as the relevant objects controlling when entanglement can appear inside subsystems under changes of reference frame.

This result led us to the classification of subsystems that admit a separable description relative to a given reference frame. Whenever a subsystem $I$ admits a separable description, QRF transformations within $I^c$ act through random local unitaries and cannot increase its internal entanglement. In this sense, separable descriptions are those that maximize the entanglement internal to a given subsystem. This is also related to the conservation law found in~\cite{Cepollaro:2024rss}: when changing reference frames, coherence in the new reference frame is converted into entanglement involving the previous frame.



The distinction between subsystems that admit a separable description and those that do not then provides a criterion for a notion of entanglement that is genuinely new within quantum reference frames. Subsystems that do not admit a separable description with respect to any complementary frame are precisely those that are entangled with their complementary systems independently of their QRF description. These are precisely the systems that pinpoint genuinely new quantum correlations within quantum reference frames.



\acknowledgements

The authors thank Leon Loveridge for carefully reading the manuscript, and Anne-Catherine de la Hamette and Jan Glowacki for insightful discussions, as well as M. Hamed Mohammady, Philipp A. H\"ohn and Carlo Cepollaro for suggestions that allowed us to contextualize our results with current literature, with special thanks to Ali Akil for suggesting a comparison with the results of~\cite{Cepollaro:2024rss}. The authors also acknowledge two anonymous referees for important comments that improved the quality of the manuscript. TRP and GF are thankful for financial support from the Olle Engkvist Foundation (no.225-0062). Nordita is partially supported by Nordforsk. NSM thanks for the funding of projects INDORBLE 09I03-03-V04-00679 for excellent researchers R2, DeQHOST APVV-22-0570, and QUAS VEGA 2/0164/25. The authors benefited from the activities of COST Action \href{https://rqi-cost.org/}{CA23115: Relativistic Quantum Information} funded by COST (European Cooperation in Science and Technology).

\appendix

\section{General Quantum Reference Frame Transformations and Subsystems}\label{app:generalPQRF}

 Consider a general state described with respect to reference frame $k\in I^c$:
\begin{equation}\label{eq:Psikgen}
    \ket{\Psi}^{(k)} = \int \dd \bm g \, \delta(g_k) f(\bm g) \ket{\bm g}_{I^c} \otimes \ket{\psi(\bm g)}_I.
\end{equation}the 
The operator that changes reference frame from $k$ to $l\in I^c$ can be written as
\begin{align} 
    \hat{T}^{k\shortto l}  = \hat{\Pi}_{kl} \int \dd g_l \openone_k&\otimes |g_l^{-1}\rangle\!\bra{g_l}_l \\&\otimes\left(\bigotimes_{i\neq k,l} \hat{U}_i^\dagger(g_l)\right) \otimes \hat{\mathcal{U}}^\dagger(g_l),\nonumber
\end{align}
where $\hat{\mathcal{U}}^\dagger(g_l) = \hat{U}_{m+1}^\dagger(g_l)...\hat{U}_N^\dagger(g_l)$ and $\hat{U}_i^\dagger(g)\ket{g_i}_i = \ket{g^{-1}g_i}_i$ is the unitary action in the reference system $\mathcal{H}_i$.
At this stage, we note that the identity in $k$ can be written as
\begin{equation}
    \openone_k = \int \dd g_k \ket{g_k}\!\!\bra{g_k}_k,
\end{equation}
so that for any operator $\hat{O}(g_l)$,
\begin{align}
    &\hat{\Pi}_{kl} \int \dd g_l \openone_k\otimes |g_l^{-1}\rangle\!\bra{g_l}_l \otimes\hat{O}(g_l)\\& =  \hat{\Pi}_{kl} \int \dd g_l \dd g_k \ket{g_k}\!\!\bra{g_k}_k \otimes |g_l^{-1}\rangle\!\bra{g_l}_l \otimes\hat{O}(g_l)
    \\& =   \int \dd g_l \dd g_k |g_l^{-1}\rangle\!\bra{g_k}_k \otimes \ket{g_k}\!\!\bra{g_l}_l \otimes\hat{O}(g_l).
\end{align}
Defining the operator
\begin{equation}
    \hat{P}_{kl}(\bm g) = \ket{e}\!\!\bra{g_k}_k \otimes \ket{g_lg_k}\!\!\bra{g_l}_l\left(\bigotimes_{i\neq k,l}\ket{g_i}\!\!\bra{g_i}_i\right),
\end{equation}
we can write the change of reference frames as
\begin{equation}
    \hat{T}^{k\shortto l}  = \int \dd \bm g \,g_l^{-1}\!\!\cdot \hat{P}_{kl}(\bm g) \otimes \hat{\mathcal{U}}^\dagger(g_l),
\end{equation}
where $g_l^{-1}\ket{\bm g} = \ket{g_l^{-1}\bm g}$ denotes the left unitary action acting on the reference systems. However, notice that this transformation should only be applied to vectors described relative to reference frame $k$ that can be written as $\ket{\Psi}^{(k)} = \ket{e}_k\otimes \ket{\Psi_\text{rest}}$. In this subspace we have $g_k = e$, and, under an integral, $\hat{P}_{kl}(\bm g) = \delta(g_k)\ket{\bm g}\!\!\bra{\bm g}$, allowing the change of reference frames to be recast simply as
\begin{equation}
    \hat{T}^{k\shortto l}  = \int \dd \bm g \,\delta(g_k)\,g_l^{-1}\!\!\cdot \ket{\bm g}\!\!\bra{\bm g} \otimes \hat{\mathcal{U}}^\dagger(g_l)
\end{equation}
when applied to vectors described with respect to reference frame $k$~\footnote{Notice that, when applying $\hat{T}^{k\shortto l}$ to a  state of the form $\ket{\Psi}^{(k)} = \ket{e}_k\otimes \ket{\Psi_\text{rest}}$, one can simply write
\begin{equation}
    \hat{T}^{k\shortto l}  = \int \dd \bm g \,g_l^{-1}\!\!\cdot \ket{\bm g}\!\!\bra{\bm g} \otimes \hat{\mathcal{U}}^\dagger(g_l),
\end{equation}
where $\delta(g_k)$ is naturally implemented by the states described with respect to frame $k$.}. 

Applying the change of reference frame transformation to the state in Eq.~\eqref{eq:Psikgen} yields
\begin{equation}
    \ket{\Psi}^{(l)} = \int \dd \bm g \, \delta(g_k) f(\bm g) \ket{g_l^{-1}\bm g}_{I^c} \otimes \hat{\mathcal{U}}^\dagger(g_l)\ket{\psi(\bm g)}_{I}.
\end{equation}
We now perform a change of variables $\tilde{g}_i = g_l^{-1}g_i$ for $i\neq l$, so that $g_i = g_l \tilde{g}_i$. By denoting $\tilde{\bm g}_{\bar{l}} = (\tilde{g}_1,...,\tilde{g}_{l-1},e,\tilde{g}_{l+1},...,\tilde{g}_m)$, we have $\bm g = g_l \bm g_{\bar{l}}$, so that the integral can be recast as
\begin{align}
    \ket{\Psi}^{(l)} = \int \dd \tilde{\bm g}_{\bar{l}}\,\dd g_l \, \delta(g_l\tilde{g}_k) & f(g_l \tilde{\bm g}_{\bar{l}}) \ket{\tilde{\bm g}_{\bar{l}}}_{I^c} \\
    &\otimes \hat{\mathcal{U}}^\dagger(g_l)|\psi(g_l \tilde{\bm g}_{\bar{l}})\rangle_{I},\nonumber
\end{align}
where integration with respect to $\tilde{\bm g}_{\bar{l}}$ denotes integration with respect to $\tilde{g}_1,...\tilde{g}_{l-1},\tilde{g}_{l+1},...,\tilde{g}_m$.

Performing the integration over $g_l$, we find $g_l = \tilde{g}_k^{-1}$ due to the Dirac delta, implying
\begin{align}
    \ket{\Psi}^{(l)} = \int \dd \tilde{\bm g}_{\bar{l}} & f(\tilde{g}_k^{-1} \tilde{\bm g}_{\bar{l}}) \ket{\tilde{\bm g}_{\bar{l}}}\otimes \hat{\mathcal{U}}^\dagger(\tilde{g}_k^{-1})|\psi(\tilde{g}_k^{-1} \tilde{\bm g}_{\bar{l}})\rangle. \label{eq:psil}
\end{align}
Finally, we notice that for any function $f(\tilde{\bm g})$ and vectors $|v(\tilde{\bm g})\rangle$,
\begin{equation}
    \int \dd \tilde{\bm g}_{\bar{l}} f(\tilde{\bm g}_{\bar{l}}) |{v(\tilde{\bm g}_{\bar{l}})}\rangle = \int \dd \tilde{\bm g} \delta(\delta{g}_l)f(\tilde{\bm g}) |{v(\tilde{\bm g})}\rangle ,
\end{equation}
allowing us to recast Eq.~\eqref{eq:psil} as
\begin{align}
    \ket{\Psi}^{(l)} = \int \dd \tilde{\bm g} & \delta(\tilde{g}_l) f(\tilde{g}_k^{-1} \tilde{\bm g}) \ket{\tilde{\bm g}}_{I^c} \otimes \hat{\mathcal{U}}^\dagger(\tilde{g}_k^{-1})|\psi(\tilde{g}_k^{-1} \tilde{\bm g})\rangle_I. 
\end{align}
Defining the states $|{\tilde{\psi}(\tilde{\bm g})}\rangle_I = \hat{\mathcal{U}}^\dagger(\tilde{g}_k^{-1})|\psi(\tilde{g}_k^{-1}\tilde{\bm g})\rangle_I$ and the function $\tilde{f}(\tilde{\bm g}) = f(\tilde{g}_k^{-1}\tilde{\bm g})$, we can then write
\begin{align}
    \ket{\Psi}^{(l)} = \int \dd \tilde{\bm g} \delta(\tilde{g}_l) \tilde{f}(\tilde{\bm g}) \ket{\tilde{\bm g}}_{I^c} \otimes|\tilde{\psi}(\tilde{\bm g})\rangle_I.
\end{align}

\section{Subsystems in Quantum Reference Frame Transformations in
Mixed States}\label{app:mixed}

A general mixed state for a reference frame $k$ and remaining systems including both $I$ and $I^c$ can be decomposed as
\begin{equation}\label{eq:rhokgen}
    \hat{\rho}^{(k)} = \int \dd \bm g \dd \bm g' \delta(g_k)\delta(g_k')f(\bm g, \bm g')\ket{\bm g}\!\!\bra{\bm g'}\otimes \hat{O}_{\bm g\bm g'},
\end{equation}
where $f(\bm g, \bm g') = f(\bm g', \bm g)^*$ and $\hat{O}_{\bm g\bm g'}$ are operators acting in the combined Hilbert space of $I$ such that $\hat{O}_{\bm g\bm g'} = \hat{O}_{\bm g'\bm g}^\dagger$ and together they satisfy $\int \dd \bm g \delta(g_k) f(\bm g, \bm g) \text{tr}(\hat{O}_{\bm g\bm g}) = 1$ due to the normalization condition for $\hat{\rho}$. Applying the change of reference frame from $k$ to $l$ then yields
\begin{align}
    &\hat{\rho}^{(l)} = \hat{T}^{k\shortto l} \hat{\rho}^{(k)} \hat{T}^{k \shortto l\dagger}\\
    &= \int \dd \bm g \dd \bm g'\delta(g_k)\delta(g_k') f(\bm g ,\bm g') \ket{g_l^{-1}\bm g}\!\!\bra{g_l'^{-1}\bm g'} \nonumber\\
    &\quad\quad\quad\quad\quad\quad\quad \quad\quad\quad\quad\quad\quad\otimes\hat{\mathcal{U}}^\dagger(g_l)\hat{O}_{\bm g \bm g'}\hat{\mathcal{U}}(g_l'),\nonumber
\end{align}
where we denoted $\hat{\mathcal{U}}^\dagger(g_l) = \hat{U}_1^\dagger(g_l)...\hat{U}^\dagger_N(g_l)$ as the total unitary action in $I$. Performing analogous computations to those in Appendix~\ref{app:generalPQRF}, we can rewrite
\begin{equation}
    \hat{\rho}^{(l)} = \int \dd \tilde{\bm g} \dd \tilde{\bm g}' \delta(\tilde{g}_l)\delta(\tilde{g}_l')\tilde{f}(\tilde{\bm g}, \tilde{\bm g}')\ket{\tilde{\bm g}}\!\!\bra{\tilde{\bm g}'}\otimes \hat{\tilde{O}}_{\tilde{\bm g}\tilde{\bm g}'},
\end{equation}
where we defined $\hat{\tilde{O}}_{\tilde{\bm g}\tilde{\bm g}'} = \hat{\mathcal{U}}^{\dagger}(\tilde{g}_k^{-1})\hat{{O}}_{\tilde{\bm g}\tilde{\bm g}'}\hat{\mathcal{U}}(\tilde{g}_k'^{-1})$ and the function $\tilde{f}(\tilde{\bm g},\tilde{\bm g}') = f(\tilde{g}_k^{-1}\tilde{\bm g},\tilde{g}_k'^{-1}\tilde{\bm g}')$. 

Notice that the partial state of $I$ for a state of the form of Eq.~\eqref{eq:rhokgen} is given by
\begin{equation}
    \hat{\rho}^{(k)}_I = \int \dd \bm g \delta(g_k) f(\bm g, \bm g) \hat{O}_{\bm g \bm g}.
\end{equation}
From this expression, together with the fact that $\hat{\tilde{O}}_{\bm g \bm g}$ are related to $\hat{O}_{\bm g \bm g}$ by the application of local unitaries, we conclude that if $\hat{O}_{\bm g \bm g}$ are separable for all $\bm g\in G^m$ in a given reference frame, then the reduced state $\hat{{\rho}}^{(l)}_{I}$ will be separable for any reference frame $l$.

\section{QRFs cannot increase entanglement between subsystems compared to their separable description}\label{app:proof}

Assume that a subsystem $I$ admits a separable description with respect to reference frame $k$, so that its reduced state is
\begin{equation}
    \hat{\rho}_I^{(k)} = \tr_{I^c}(\ket{\Psi}\!\!\bra{\Psi}^{(k)}) = \ket{\psi}\!\!\bra{\psi}_I^{(k)} 
\end{equation}
Then the effective quantum channel acting on $I$ after a quantum reference frame transformation is of the form
\begin{align}
    &\hat{\rho}_I^{(l)} = \mathcal{E}_{k\shortto l}(\ket{\Psi_I}\!\!\bra{\Psi_I}^{(k)}) \\
    &= \int \dd \bm g \delta(g_l)|\tilde{f}(\bm g)|^2 \hat{\mathcal{U}}^\dagger(g_k^{-1})\ket{\psi_I}\!\!\bra{\psi_I}^{(k)} \hat{\mathcal{U}}(g_k^{-1}),\nonumber
\end{align}
where $\hat{\mathcal{U}}^\dagger(g_k^{-1})$ are local unitaries to each system. This map is then a random local unitary map, and cannot increase entanglement. Notice that although one can invert the QRF transformation mapping $\hat{\rho}_I^{(l)}$ to $\hat{\rho}_{I}^{(k)} = \ket{\Psi_I}\!\!\bra{\Psi_I}^{(k)}$, the channel $\mathcal{E}_{k\shortto l}$ is only invertible when it is a unitary channel. That is, if $f(\bm g)$ is only non-zero for a given value of $g_k = g_{k,0}$. This would only happen if the $l$ itself provided a separable description for system $I$, in which case the entanglement in the states $\ket{\psi_I}^{(k)}$ and $\ket{\psi_I}^{(l)}$ would be the same, due to the local character of the unitaries $\hat{\mathcal{U}}^\dagger(g_k^{-1})$. In general, a change of QRF can only be thought of as a channel acting on $I$ when going from a separable description to another reference frame. The explicit form of this channel then allows us to conclude that separable descriptions of a system are the ones that contain the most entanglement within $I$.

\bibliography{references}

\end{document}